# Simulation study of ballistic spin-MOSFET devices with ferromagnetic channels based on some Heusler and oxide compounds

Patrizio Graziosi[1*] and Neophytos Neophytou[2]

[1]CNR – ISMN, via Gobetti 101, 40129, Bologna, Italy

[2]School of Engineering, University of Warwick, Coventry, CV4 7AL, UK

[*] patrizio.graziosi@gmail.com

Newly emerged materials from the family of Heuslers and complex oxides exhibit finite bandgaps and ferromagnetic behavior with Curie temperatures much higher than even room temperature. In this work, using the semiclassical top-of-the-barrier FET model, we explore the operation of a spin-MOSFET that utilizes such ferromagnetic semiconductors as channel materials, in addition to ferromagnetic source/drain contacts. Such a device could retain the spin polarization of injected electrons in the channel, the loss of which limits the operation of traditional spin transistors with non-ferromagnetic channels. We examine the operation of four material systems that are currently considered some of the most prominent known ferromagnetic semiconductors, three Heusler-type alloys ($Mn_2CoAl$, CrVZrAl, CoVZrAl) and one from the oxide family ($NiFe_2O_4$). We describe their bandstructures by using data from DFT calculations. We investigate under which conditions high spin polarization and significant $I_{ON}/I_{OFF}$ ratio, two essential requirements for the spin-MOSFET operation, are both achieved. We show that these particular Heusler channels, in their bulk form, do not have adequate bandgap to provide high $I_{ON}/I_{OFF}$ ratios, and have small magnetoconductance compared to state-of-the-art devices. However, with confinement into ultra-narrow sizes down to a few nanometers, and by engineering their spin dependent contact resistances, they could prove promising channel materials for the realization of spin-MOSFET transistor devices that offer combined logic and memory functionalities. Although the main compounds of interest in this paper are $Mn_2CoAl$, CrVZrAl, CoVZrAl, and $NiFe_2O_4$ alone, we expect that the insight we provide is relevant to other classes of such materials as well.

# I. Introduction

Low power operation, storage and computing functionalities embedded in the same device, are among the advantages of the spin-MOSFET, the charge-based beyond-CMOS prime transistor device candidate.[1-4] Current spin-MOSFETs have ferromagnetic source and drain contacts, but a non-magnetic channel (such as Silicon). The current-voltage characteristics are controlled by the gate voltage and by the relative magnetization orientations of the source and drain, which act as the spin injector and the spin detector, respectively. Thus, spin-MOSFET devices can be used as both, a logic transistor and a memory storage element since the parallel (P) and the antiparallel (AP) magnetization directions of the source and drain result in different resistive behaviors. The state-of-the-art spin-MOSFET proposed by Toshiba[4-6] consists of ferromagnetic source and drain contacts connected through tunnel barriers to heavily doped silicon regions to overcome the "conductivity mismatch" issue.[7] The 'write' operation is performed with a magnetic tunnel junction (MTJ) that sets the magnetization of the drain via the spin transfer torque (STT) effect,[5, 6] while the magnetization direction of the source is kept unchanged.

Despite the large efforts over the last several years, a complete experimental demonstration of the spin-MOSFET device has not yet been achieved, except at very low temperatures (12 K)[5, 6] however, reliable operation up to 400 K is required.[8] One reason for this failure is the loss of the current spin polarization due to spin scattering in the non-ferromagnetic channel.[9, 10]

A channel composed of a ferromagnetic semiconductor (FS), which effectively transfers the spin information from source to drain, could retain the spin polarization. FSs can be achieved by magnetic doping (as in the case of "diluted magnetic semiconductors" – DMS [11, 12]), but usually $T_C$ is less than 200 K, which forbids their use in electronic devices.[8, 13] Very high $T_C$ (>700K) can be measured in bulk DMS (due to segregated ferromagnetic clusters, for instance), but in that case there is no separation between majority and minority spin bands,[11, 13] while it is imperative that the ferromagnetism originates from the material's bandstructure and is not a result of spurious effects.[11, 13] Recent developments in materials science, however, have demonstrated semiconductor compounds that are actually intrinsically ferromagnetic (not due to doping) with $T_C$ > 400

K. Such compounds can be traced in the Heusler [14-18] and the oxide families.[19, 20] Thus, in this work, we computationally explore a spin-MOSFET device in which the channel is composed of recently demonstrated FSs from the Heusler families, more precisely the Heusler alloys with Y-type or XA-type lattice (in particular alloys CrVZrAl, CoVZrAl and $Mn_2CoAl$) and oxide families (in particular $NiFe_2O_4$), and elaborate on the conditions that will make such materials suitable for logic and memory spin-MOSFET applications. The specific materials we consider are described using DFT bandstructure extracted band offsets and effective masses. We explore bandstructure conditions to achieve the highest spin polarization (SP) in the channel and the conditions for both high SP and $I_{ON}/I_{OFF}$ ratio, as well as how quantum confinement and spin dependent contact resistances can be engineered to improve the device performance.

## II. Approach

We consider a symmetric device in which the source and drain are identical in order to account for an easier fabrication process (although in principle the source can be non-magnetic). The source and the ferromagnetic channel are always aligned in parallel, while the ferromagnetic drain is switched via STT-MTJ, as in Toshiba's approach[4-6] (see **Fig. 1a**). This is somewhat similar to a Schottky barrier MOSFET previously proposed,[21] but here we consider Ohmic contacts and parameters from real material bandstructures.

The semiclassical top-of-the-barrier ballistic model (FETToy[22]), validated in the past for various other materials,[23, 24] is used to simulate the transistor behavior including self-consistent electrostatics.[24, 25] We assume a 1.1 nm $SiO_2$ gate oxide. The model assumes that the positive going states are filled according to the source Fermi level $E_{FS}$, whereas the negative going states according to the drain Fermi level, $E_{FD}$. The current, within the Landauer formalism, is the difference of the two fluxes.[25]

We describe the material using multiple majority and minority spin bands, both for valence (VB) and conduction bands (CB) combined, contributing to positive and negative charges. Thus, a charge neutrality level is set in the simulation to begin with, for the initial position of the Fermi level $E_F$. $E_F$ is set to -0.1 eV arbitrarily, which only affects the shift in the threshold voltage of the channel. Room temperature $T$ = 300 K is considered. To keep the bandstructure features qualitatively simple, we assume 1D, parabolic, isotropic bands, despite the fact that the bands can be non-parabolic and anisotropic. However, the effective masses, band degeneracies, and band splittings for each compound we consider are extracted from DFT data that are presented in various references in existing literature and summarized in **Table 1**. [14, 17-19] We consider only the bands around the Fermi level, which are more involved in transport (lower CBs and higher VBs), in the directions Γ−X for $Mn_2CoAl$ and CoVZrAl and Γ−K for CrVZrAl. In the case of the oxide $NiFe_2O_4$, we adopt an average conductivity effective mass $\frac{1}{m_a} = \frac{1}{2m_\mathrm{R}} + \frac{1}{2m_\mathrm{T}}$, following the approach commonly used in semiconductors,[26] where $m_\mathrm{R}$ and $m_\mathrm{T}$ are the effective masses along Γ−R and Γ−T respectively.[27]

Thus, our approach provides useful first order guidance into the effect of the bandstructure features in electronic transport and spin-MOSFET device operation of such materials. The spin-polarization (SP) of the current is defined as $\mathrm{SP} = \left(\mathrm{I}^{\uparrow} - \mathrm{I}^{\downarrow}\right)/\left(\mathrm{I}^{\uparrow} + \mathrm{I}^{\downarrow}\right)$ where $\mathrm{I}^{\uparrow(\downarrow)}$ is the majority (minority) spin currents in the channel. The channel is assumed to be coupled to spin dependent source/drain resistances as indicated later on in **Fig. 3a**, and we then compute the device current in a post-processing step using a bi-dimensional linear interpolation scheme.[28] In this scheme, the bi-dimensional current matrix for each spin orientation separately is mapped as $I(V_\mathrm{D}, V_\mathrm{G}) \coloneqq I(V_\mathrm{D}', V_\mathrm{G}')$ where $V_\mathrm{D}' = I(V_\mathrm{D})R_\mathrm{S} + V_\mathrm{D}$, $V_\mathrm{G}' = I(V_\mathrm{G})R_\mathrm{S} + V_\mathrm{G}$, and $R_\mathrm{S}$ is the total series resistance coming from the sum of the source and drain contact resistances. Then, the $I(V_\mathrm{D}', V_\mathrm{G}')$ two-dimensional current data matrix is linearly interpolated on the original ($V_\mathrm{D}$, $V_\mathrm{G}$) set.

# III. Results and Discussion

**Figure 1b** shows a generic bandstructure used to investigate what parameters lead to high SP necessary for memory functionality, and high $I_{ON}/I_{OFF}$ ratio necessary for computation functionality. Majority (minority) spin bands are shown in **blue (red)**. The three basic parameters we consider are the energy gap of the majority spins $E_G^{\uparrow}$, the splitting between the two spin bands Δ, and the bands' effective masses $m^{\uparrow(\downarrow)}$.

For this first order evaluation of SP, in **Fig. 1c** we use near equilibrium conditions with low drain bias $V_D$ = 0.1 mV and effective masses of $m^{\uparrow(\downarrow)} = m_0$, where $m_0$ is the rest mass of the electron. At this point, we neglect the series resistances. We choose band energy values as noted in the figure, typical for Heuslers like CoVZrAl, CoVTiAl, and $Mn_2CoAl$ (see also **Table 1**).[14, 17, 18] **Figure 1c** summarizes the SP dependence of the materials with different bandstructure parameters. Starting from the bandstructure we show in **Fig. 1b** (**blue line** for SP in **Fig. 1c**) we observe that by increasing Δ, the SP rises (to the **yellow line**). SP is also retained for higher gate biases as well, since the bands of the minority carriers reside at higher energies and so they have a smaller occupancy. Decreasing $E_G^{\uparrow}$ until the majority CB and VB overlap (noted by negative bandgap values in **Fig. 1c**) as in the case of $Mn_2CoAl$, further improves slightly the SP for $V_G$ higher than 0.7 V, because this increases the majority DOS contribution (**green dashed-dotted line**). Increasing the majority effective mass (by 3×), further improves the SP (**red-dashed line**). A heavier majority band allows much less shift in the bands with $V_G$, thus the $E_F$ remains within the majority band. Thus, in order to have high SP in ballistic channels, we seek high Δ, large effective mass for the majority spins (in the case of scattering dominated channels larger masses would of course induce more scattering and this condition needs to be re-examined), and in general a small $E_G^{\uparrow}$.

On the other hand, a small $E_G^{\uparrow}$ reduces the $I_{ON}/I_{OFF}$ ratio. To have a ratio $I_{ON}/I_{OFF} \sim 10^3$ evaluated at $V_D$ = 0.75 V, the ITRS (International Technology Roadmap for Semiconductors) specified voltage for the 2020 technology node,[29] the gap should be at least 1.1 eV (similar to the Si gap). To explore the effect of the $E_G^{\uparrow}$ on the SP at the specified voltage $V_D$ = 0.75 V, we set the values of Δ to 0.3 eV and of $m^{\uparrow(\downarrow)}$ to $m_0$ and vary $E_G^{\uparrow}$ from the overlap condition ($E_G^{\uparrow}$ = -0.1eV) that maximizes SP in the low bias regime, to 1.1 eV, that is required for sufficient $I_{ON}/I_{OFF}$ ratio. **Figure 1d** reports the results for three bandgap values at $V_D$ = 0.75 V: $E_G^{\uparrow}$ = -0.1 eV, typical also of spin gapless Heuslers like bulk $Mn_2CoAl$, $E_G^{\uparrow}$ = 0.3 eV, representative of Heuslers like bulk CoVZrAl, and $E_G^{\uparrow}$ = 1.1 eV. We chose Δ = 0.3 eV and $m^{\uparrow(\downarrow)} = m_0$ for all cases. In this high drain bias case, band overlap results in reduced spin polarization. The relatively high drain bias sets $E_{FD}$ too low, closer to the minority VB, which thus begins to contribute to transport. A higher $E_G^{\uparrow}$ value prevents this and still allows a high SP at the required drain bias. The spin gapless semiconductors with a zero, or negative, bandgap for majority spins and a finite bandgap for the minority, have almost 100% SP only at very low drain biases. Thus, they can act as very efficient spin injecting electrodes but not spin-polarized channels for spin-MOSFETs.

To consider both the $I_{ON}/I_{OFF}$ ratio and SP in device performance evaluation, we introduce here a performance indicator $Q$, defined as the product of the highest SP($V_G$) and the highest $I_{ON}/I_{OFF}(V_G)$, both of them evaluated at $V_D = 0.75$ V, and both being gate voltage dependent. These parameters are evaluated at different $V_G$ because they correspond to different functionalities ($I_{ON}/I_{OFF}$ for logic and SP for memory). **Figure 1e** plots the $Q$ values for the four illustrative bandstructure parameters. $Q$ should be at least in the $10^3$-$10^4$ range to ensure sufficient operation. By ignoring the majority/minority band separation in (i), $Q$ drops to zero because SP is zero. Increasing Δ to 0.1eV in (ii) increases SP to ~0.5 but the $I_{ON}/I_{OFF}$ ratio is too low to gain a decent $Q$. In (iii) where $E_G^{\uparrow}$ is increased to 1.1 eV (to obtain a good $I_{ON}/I_{OFF}$ ratio) and the split between the majority and minority spin bands to 0.3 eV (typical of many Heuslers) we achieve an acceptable $Q$ with SP~1. Further improvement is observed by decreasing the electron effective mass to $0.1m_0$ for both the spin orientations because of higher $I_{ON}/I_{OFF}$ ratio (iv) – i.e. when comparing at the same $V_G$ the bands are shifted further towards the Fermi level when the effective mass is lower, which raises $I_{ON}$.

Because common bulk ferromagnetic Heuslers do not usually possess the essential bandgap to provide large $I_{ON}/I_{OFF}$ ratios, one way to increase the bandgap is to use ultra-narrow Heusler channels where the bandgap is increased due to quantum confinement. A simple estimate of the bandgap increase can be provided by the particle-in-a-box quantization theory. The energy shift of the bands is calculated as $E_n = \frac{\pi^2 \hbar^2}{2mt^2}$ where $t$ is the material confinement size, and $m$ is the effective mass, which is assumed to be isotropic here. One needs to be aware that due to the complexities of the materials' bandstructures this approach could only serve as a crude indication. In fact, there are no adequate theoretical or experimental studies that investigate confinement effects of Heusler compounds to-date from which we could have extracted more accurate information.[30, 31] Nevertheless, to obtain a more realistic bandgap behavior with confinement, we estimate the band edges using the approach described in Ref. [23, 24]. In this approach, the confined band edges are approximated by the actual bulk DFT non-parabolic band values at an equivalent quantized wavevector of value $k_L=\pi/t$ where $t$ is the confinement length. [24]

The Heusler compound CrVZrAl, for example, is a ferromagnetic semiconductor with $E_G^{\uparrow}$ ~ 0.66 eV (**Fig. 2a**) and can only allow a small $I_{ON}/I_{OFF}$ ratio. **Figure 2b** shows the shift of the band edges with confinement dimension (thickness of a quantum well or diameter of a quantum wire labelled here as *t*). Because of the relatively high masses (in the 0.4 – 3.0 $m_0$ range, see **Table 1**) it is necessary to reduce the confinement dimension below $t$ = 3 nm to achieve a sizeable bandgap increase. In this case, the $Q$ factor is largely improved for the ultra-narrow Heusler channel compared to bulk, whereas the SP is also close to one at the considered biases (**Fig. 2c**). The dotted lines in **Fig. 2b** and **Fig. 2c** indicate the band edges and $Q$ factor for non-parabolic band considerations. Non-parabolicity slows down the bandgap increase for thicknesses below $t$ < 2.5 nm, similarly to the case of Si,[32] which reflects in smaller $Q$ values.

Due to the large number and widely varying properties of Heusler compounds, each alloy can have distinct behavior. **Figure 2d** shows indicatively the bulk $Mn_2CoAl$ bandstructure, which is a widely studied spin gapless semiconductor with a finite band gap of ~ 0.5 eV for the minority spin (**red lines**), but a zero band gap (a small overlap of 0.03 eV in fact) for the majority spin (**blue lines**). Due to the zero bandgap, transistor operation is prohibited, but confinement in ultra-narrow dimensions would provide a finite bandgap. **Figure 2e** shows how the band edges move with increasing confinement. For this Heusler, the layer thickness has to be reduced below $t$ < 1.5 nm to achieve sufficient $Q$ values as shown in **Fig. 2f**. Its somewhat lower conduction band effective mass compared to CrVZrAl (see **Table 1**), results in larger shift of the band edge, but the smaller bulk bandgap requires further thickness scaling. Considering non-parabolocity effects, as depicted by the dotted lines in **Fig. 2e** and **Fig. 2f**, shows that this material needs to be scaled to unrealistically small sizes to be a useful transistor channel material.

We note here that we assumed that the properties of these materials behave according to the conventional particle-in-a-box confinement trend. In the case of the another Heusler compound, the $Co_2MnSi$, thin films down to 70 nm have been demonstrated to maintain the half metallic bulk behavior.[33] For smaller thicknesses, studies indicate that the bulk magnetic properties are maintained until 10 nm (although the half metallicity was not verified), but for lower thicknesses the magnetic properties gradually deteriorate.[34] Such results, however, should not prevent the employment of Heusler alloys within the proposed device concept. Ferromagnetic semiconducting Heuslers, also named 'spin filters', have been studied only recently.[14, 18] There is a large number of unexplored Heusler compounds, which means that the search for ferromagnetic Heusler semiconductors with larger $E_G^{\uparrow}$ and possibly lower masses ($\leq 0.1m_0$) is required and *timely*. New materials are likely to be identified, for which $E_G$, $\Delta$, confinement behaviour, strain behaviour, etc., would be addressed for each material separately.[15, 16, 35]

We now describe the operation of the actual spin-MOSFET device. As a bandstructure example we adopt the one from **Fig. 1d** with $m_h = 1m_0$ and $m_e = 0.1m_0$, $E_G^{\uparrow} = 1.1$ eV and $\Delta = 0.3$ eV (typical values that could provide proper operation). The spin-MOSFET has spin dependent series resistances (different for majority and minority spins) at the source/drain contacts. These contact resistances arise from the fact that even in an ideal contact between two ferromagnetic materials, spin flip events occur at the interface because of the different spin resolved density of states and group velocities in the junction materials.[36] Thus, even in the absence of any 'traditional' contact resistance, the interfacial spin scattering introduces a junction resistance. **Figure 3a** shows the device model with the resistances used for the simulations.

The contact resistances are estimated from a model for ideal contacts where an interfacial voltage drop $\Delta V_I$ occurs for spin flip events. The model is detailed in Ref. [36] and in the **Appendix**, and summarized as follows:[36] Consider a junction between two ferromagnetic materials A and B (in Ref. [36] the two materials A and B are considered to be the same, while in our case they are different). The interfacial voltage drops for the parallel (P) and antiparallel (AP) cases at the junction of these materials are $\Delta V_I^{\mathrm{P}} = \frac{J}{2}(\mathrm{SP_A}\Delta\mathrm{SP}\rho^{\mathrm{A}} l_s^{\mathrm{A}} + \mathrm{SP_B}\Delta\mathrm{SP}\rho^{\mathrm{B}} l_s^{\mathrm{B}})$ and $\Delta V_I^{\mathrm{AP}} = J(\mathrm{SP_A^2}\rho^{\mathrm{A}} l_s^{\mathrm{A}} + \mathrm{SP_B^2}\rho^{\mathrm{B}} l_s^{\mathrm{B}})$ respectively, where $J$ is the current density, $\rho_{A(B)}$ the resistivity of the material, $l_s$ the spin diffusion length inside the ferromagnet, and $\Delta\mathrm{SP} = \mathrm{SP_A} - \mathrm{SP_B} > 0$ (see **Appendix**). We assume A is a Heusler alloy with $SP_A = 0.95$, $\rho_A = 1\cdot10^{-4}\ \Omega$cm [16] and $l_s^A = 3$ nm,[37] and B a Permalloy with $SP_B = 0.45$,[38] $\rho_B = 1.5\cdot10^{-5}$ Ωcm and $l_s^B = 5.5$ nm.[39] We then obtain $\Delta V_I^{AP}/\Delta V_I^{P} \sim 4$. In order to consider more detrimental events that could exist at a Heusler interface,[40] we lower this value by ~ 40% to 2.5, since the loss of SP due to defects could range from 20 to 45%.[40] Thus, $\Delta V_I/J$ ~ $10^{-11}$ $\Omega m^2$, so for a 100 $nm^2$ area junction the resistance is $R_\uparrow \sim 10^5$ Ω. The relative values for the cases we examine are denoted in **Fig. 3a**.

From a practical point of view, it is possible that a thin non-magnetic layer (~ 1 nm) between the ferromagnetic drain and the ferromagnetic channel is necessary to decouple their magnetizations.[41] This could be metallic, for instance Al in the case of $Mn_2CoAl$, or insulating, and could even increase the contact resistances (thus increasing the magnetoconductance MC if the spin dependent resistances dominate more), but does not alter the basic concept, and therefore we neglect it here.

**Figure 3b** shows in **black lines** the $I_D$ versus $V_G$ for $V_D = 0.75$ V for the P (**solid**) and AP (**dash-dot**) configurations for the calculated resistances $R_\uparrow = 10^5\,\Omega$ and $R_\downarrow = 2.5\times10^5\,\Omega$. The vertical **blue solid lines** represent the ON and OFF gate biases that approximately provide the maximum achievable $I_{ON}/I_{OFF}$ ratio for a $V_G$ window of 0.75 V.[29] The corresponding ratio is around $I_{ON}/I_{OFF} \sim 10^4$. The vertical **blue dotted line** represents the 'read' bias, which is the $V_G$ of the maximum magnetoconductance MC = $(I_P-I_{AP})/I_{AP}$. Thus, a memory bit is 'written' by defining the magnetization orientation of the drain as in Toshiba's concept,[6] and 'read' as a MC. That is, if the drain spin polarization is parallel to the current spin polarization, the current value is ~1.6 μA and the bit is read as '1', whereas in the antiparallel case the current value is ~1 μA and the bit is read as '0'. Thus, the device operation is characterized by three gate bias values: $V_G^{on}$, $V_G^{off}$ and $V_G^{read}$. The former two are used for logic computation in both the P and AP states, while the latter is used to read the stored bit.

**Figure 3c** shows by the **black line** the MC for the bandstructure of the Heusler example we consider. The MC for this reference example has a maximum ~ 60%, which is, however, lower compared to state-of-the-art devices, which is in the range of 150 – 300 %.[42] A way to increase this to higher values is to use larger spin splitting Δ, but more importantly change the ratio between the majority and minority contact resistances. Indeed, **Fig. 3b** shows in **red lines** the drain current for a case where the minority spin resistance is larger by an order of magnitude at $R_{\downarrow} = 10^6\ \Omega$. This reduces the contact resistance ratio by 4× to $R_{\uparrow}/R_{\downarrow} = 0.1$. This ratio is different from what we have calculated in the model of ideal contacts above,[36] however, this is not unrealistic for tailored interfaces and contact engineering, for instance, by introducing spin dependent tunnel barriers like MgO in iron based contacts.[43] In fact, $Co_2Cr_{0.6}Fe_{0.4}Al$/MgO/CoFe junctions have shown a resistance-area product of some $k\Omega/\mu m^2$, [44] in the same order as the contact resistances we use (we assume a 100 $nm^2$ contact area). In such case the ON conductance slightly decreases, the $I_{ON}/I_{OFF}$ ratio does not suffer too much (retains the same order of magnitude), the 'read' current drops to ~0.4 μA, but the MC strongly improves to 300% as shown by the **red line** in **Fig. 3c**. By the **green-dashed line** we show the change in MC as the spin splitting alone is doubled from $\Delta = 0.3$ eV to $\Delta = 0.6$ eV. Doubling the spin splitting $\Delta$ allows the MC to reach only slightly higher value at higher gate biases, and therefore by itself is not enough to improve the MC.

We note that although present memory devices for Magnetic Random Access Memory (MRAM) demand MC at 150 – 300 %,[42] the early spin valve devices that enabled the impressive increase in storage density in the 90s had a MC well below 20% .[45-47] Thus, the MC we calculate for $R_{\uparrow}/R_{\downarrow} = 0.4$ and $\Delta = 0.3$ eV (**black lines**) can still be applicable for memory functionality, and this lower value could be compensated by the possibility of having a non-volatile memory embedded in the CPU,[48] or by the allowed reconfigurable logic functionality.[4] However, due to the large number of Heusler alloy possibilities, other compounds could be identified in the future, or the contacts could be accordingly engineered, to provide more favorable spin dependent contact resistances. For instance, the use of ferromagnetic metallic Heusler alloys for source and drain instead of Permalloy, could lead to a much more improved spin dependent resistances ratio because the source/drain would have a higher SP.[49]

Finally, we examine the spin-MOSFET device operation in channels with realistic material bandstructures for the $Mn_2CoAl$, the CrVZrAl, the CoVZrAl, and the $NiFe_2O_4$ oxide (**Fig. 4**). As discussed above, to increase the bandgap of the Heuslers we consider confined channels ($t = 1.5$ nm for $Mn_2CoAl$, $t = 2$ nm for CrVZrAl, and $t = 1.5$ nm for CoVZrAl), but still consider the bulk $NiFe_2O_4$. The approximate bandstructures are shown in the insets of **Fig. 4a**, **4c**, **4e**, and **4g**, respectively. The effective masses and band offsets are as explained extracted from bulk DFT data as shown in **Table 1**.[14, 17-19]

**Figure 4a** shows the drain current versus $V_G$ for the P and AP states at $V_D = 0.75$ V for the $t = 1.5$ nm thick $Mn_2CoAl$ based device. In this case, quantum confinement strongly moves the lowest majority CB, but less the heavier mass lowest minority band. Thus, this ultra-confined material acquires a large bandgap (> 1 eV), and the $I_{ON}/I_{OFF}$ ratios in **Fig. 4a** are as high as $10^3$ in both the P and AP states, with $V_G^{off} = 0.2$ V and $V_G^{on} = 0.95$ V. The MC, however, plotted in **Figure 4b** at $V_D = 0.75$ V, features low values below 20%. On the other hand, the introduction of the non-parabolicity correction, as shown by the dotted line in **Fig. 4b**, increases the MC values to ~ 40% in most of the 'ON' bias region. Non-parabolicity, as shown above in **Fig. 2e**, weakens the band shift so that the band gap is smaller, but the separation Δ between the lowest majority and minority CBs becomes higher as a result of the different energy shift of the bands. This causes higher MC, but at expense of lower $I_{ON}/I_{OFF}$ ratio (not shown).

The zero bandgap disadvantage of $Mn_2CoAl$ resulted in scaling at $t = 1.5$ nm, which could be technologically challenging. We further consider two more Heusler compounds as spin-MOSFET channels with finite bandgap in their bulk form, CrVZrAl ($E_G = 0.66$ eV) and CoVZrAl ($E_G = 0.25$ eV). These channels still need to be confined to acquire useful bandgap. Unfortunately, their higher masses (compared with $Mn_2CoAl$) still require strong scaling to $t \leq 2$ nm channel thicknesses. The I-V characteristics and MC for CrVZrAl are shown in **Fig. 4c-d**. The $I_{ON}/I_{OFF}$ ratios in **Fig. 4c** are as high as $10^3$ in both the P and AP states respectively, with $V_G^{off} = 0.7$ V and $V_G^{on} = 1.45$ V. The MC, plotted in **Figure 4d** at $V_D = 0.75$ V, features values of ~ 40% around $V_G^{on}$. Similarly, for CoVZrAl in **Fig. 4e-f** the $I_{ON}/I_{OFF}$ ratios (**Fig. 4e**) are as high as $10^4$ in both the P and AP states respectively, with $V_G^{off} = 0.2$ V and $V_G^{on} = 0.95$ V. The MC, in **Figure 4f** at $V_D = 0.75$ V, features values of ~ 50% in the $V_G$ region between 0.3 V and 1.4 V. Thus, the main observation here is that the bulk $E_G^{\uparrow}$ appears to be the limiting factor for these ferromagnetic semiconducting Heuslers to be employed as spin-MOSFETs, although in the future other more suitable compounds with larger bandgaps could be identified.

In **Fig. 4g-h** we show the current-voltage characteristics and MC, respectively, for a spin-MOSFET device with bulk $NiFe_2O_4$ as channel material. The series resistances we use are the same as the ones deduced for the magnetic Heusler compounds, as such values are available for only a very few Heusler-type alloys. (To extract the actual resistances only $\rho$ is known for $NiFe_2O_4$. To the best of our knowledge, there is no information about $\beta$ and $l_s$). In this case the $I_{ON}/I_{OFF}$ ratios are ~ $10^6$ for both the P and AP states respectively (**Fig. 4g**), measured by scanning the $I_D$-$V_G$ characteristics with a 0.75V $V_G$ window, with $V_G^{off}$ at 1.4 V and $V_G^{on}$ at 2.15 V. The AP current, however, is higher compared to the P current, which leads to negative MC of ~ -70% (**Fig. 4h**). Such negative MC comes from the crossing of the majority and minority bands that makes the AP current higher than the P current. It is not rare to find spintronic devices that show negative MC – a ferromagnetic material (metal or semiconductor), does not necessarily inject majority spin electrons. Most of the occupied states below the Fermi level are aligned as the macroscopic material and are named 'majority' (the others are named 'minority'), but only the electrons closest to Fermi contribute to transport. Those could be of 'minority' spin configuration, i.e. populated with electrons aligned antiparallel to the total spin polarization of the material, as in Cobalt and Nickel.[50] This would provide negative MC.[51] Hence, when the drain and the channel are aligned in the antiparallel configuration (compared to the source magnetization), higher conductance is achieved, compared to parallel alignment.

Finally, we note that here we explored only four materials for the performance of the spin-MOSFET, however, several other materials can be identified.[52] For instance, $Ti_2VSb$ Heusler family indicates promising properties,[53] and Cr based alloys seem to have higher bandgaps.[18, 54] In addition, very recently ferromagnetic semiconductors have been predicted in the 2D layered Iron hydroxide with bandgap of around 0.65 eV. [55] While ferromagnetic metallic Heuslers ($L2_1$ lattice) have a long history [30] and the research on "Inverse-Heuslers" (XA lattice) compounds is taking off, [56, 57] ferromagnetic semiconducting Heuslers that belong to the quaternary Heusler family too, are a rather recent and perhaps underestimated subject of study.[14, 18, 58] Thus, in this work we aim at enlightening the relevance and timeliness of this materials research direction.

# IV. Conclusions

In summary, we have explored the possibility of spin-MOSFET devices with recently realized ferromagnetic semiconductor channels based on Heusler and oxide compounds. This approach transfers the magnetic functionality from the source to the channel and preserves the spin information more effectively compared to current spin-MOSFET devices. Among the multiple parameters needed for a proper spin-MOSFET device we have examined the $I_{ON}/I_{OFF}$ ratio, SP, and MC based on realistic bandstructure features taken from actual materials as either bulk or confined channels. We show that these materials could enable a new kind of spin-MOSFET with a spin-polarized channel rather than only spin-polarized contacts once confinement is utilized to improve their bandgap, and/or contact engineering provides significantly different resistances for each spin channel, which increases magnetoconductance. Importantly, Heusler thin films can be deposited in silicon compatible processes,[15, 16] even in the most complex quaternary alloys,[35, 59] which increases their technological appeal. The proposed device could be a promising candidate for the realization of spin-MOSFETs with room temperature operation and large spin polarization robustness that combine logic and memory functionalities. It could provide the advantage of having only one component for processing and data storing, reducing the number of components, the time transfer between the processing unit and the memory unit (RAM and HDD/SSD), and the parasitic capacitance related to the interconnects, all of which are targets set by the ITRS for beyond-CMOS charge-based-devices.[1]

Acknowledgements: PG has received funding from the Italian government (PRIN project No. 2015HYFSRT). NN has received funding from the European Research Council (ERC) under the European Union’s Horizon 2020 Research and Innovation Programme (Grant Agreement No. 678763). We thank Gerhard Fecher (Max-Planck-Institute, Institute for Chemical Physics of Solids) for the information about the $Mn_2CoAl$ bandstructure, Murat Tas (University of Patras, School of Natural Sciences) for the information about the CrVZrAl bandstructure, Markus Meinert (Bielefeld University, Faculty of Physics) and Francesco Mezzadri (University of Parma, Department of Chemistry) for the information about the $NiFe_2O_4$ unit cell symmetry.

Figure 1:

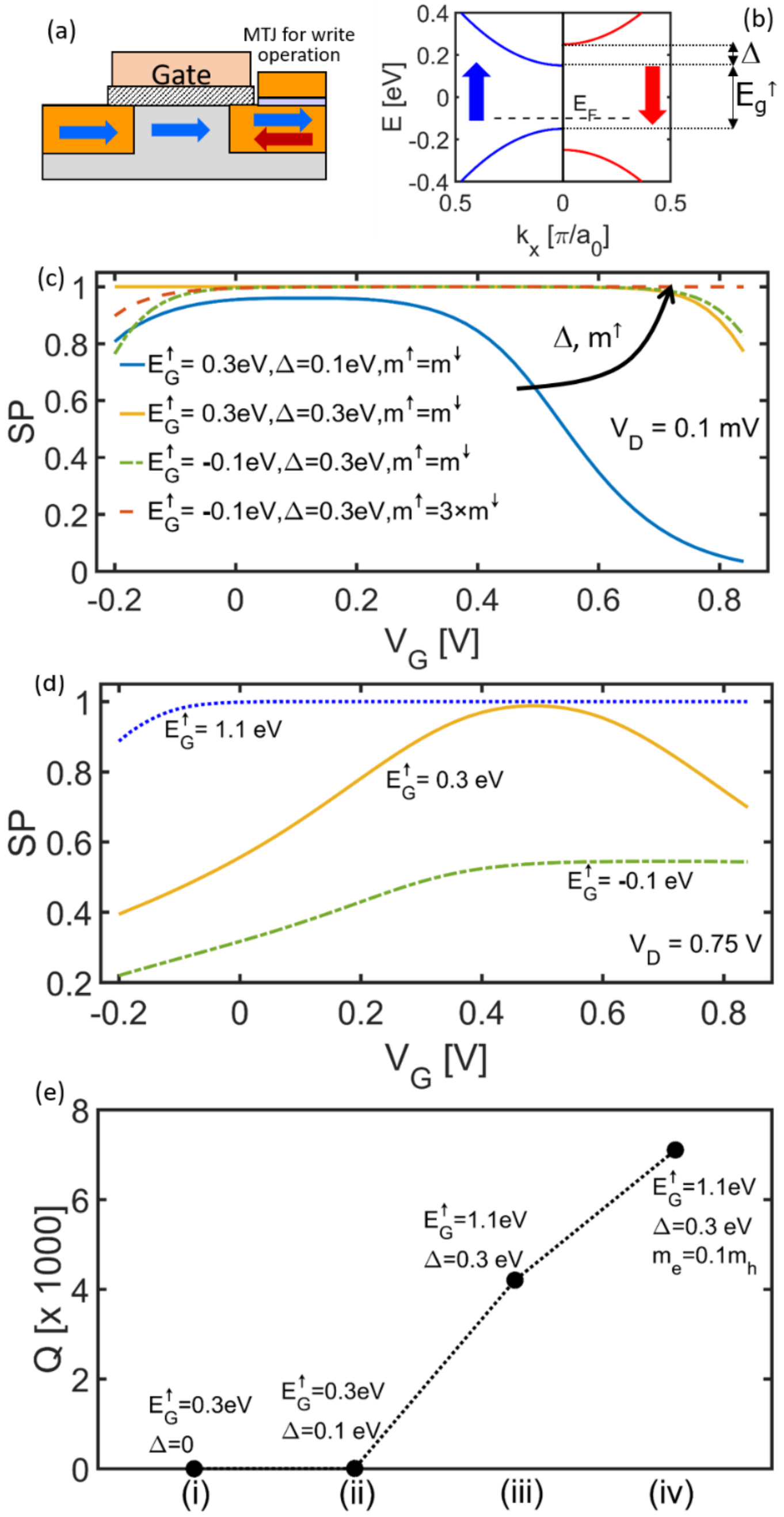

## Figure 1 caption:

(a) Device schematic with the arrows indicating the magnetization direction. The drain magnetization is switched (blue to red and reversely) by using an MTJ that exploits the STT effect as in the 'Toshiba' device concept.[4] The direction of the blue arrow in the drain represents the parallel (P) configuration while the direction of the red arrow represents the antiparallel (AP) configuration. (b) Generic bandstructure featuring majority (blue) and minority (red) bands. (c) Spin polarization (SP) versus gate bias $V_G$ for $V_D$ = 0.1 mV (low-bias conditions) for four different material bandstructures as the spin-MOSFET channels. The parameters (bandgap, spin band splitting and effective mass) of the different bandstructures are noted. (d) Spin polarization (SP) versus gate bias $V_G$ for $V_D$ = 0.75 V for three different material bandstructures as the spin-MOSFET channels. The other bandstructure parameters are kept constant at $\Delta$ = 0.3 eV and $m^{\uparrow(\downarrow)} = m_0$ while the bandgap is varied as noted. The bands overlap case (negative $E_G^{\uparrow}$ = -0.1 eV), that corresponds to a spin gapless semiconductor, gives the best SP in the low bias regime but its performance drops at practical $V_D$. (e) Performance factor $Q$ defined as the product between the $I_{on}/I_{off}$ ratio and the highest SP for $V_D$ = 0.75 V for four bandstructures with parameters as indicated.

Figure 2:

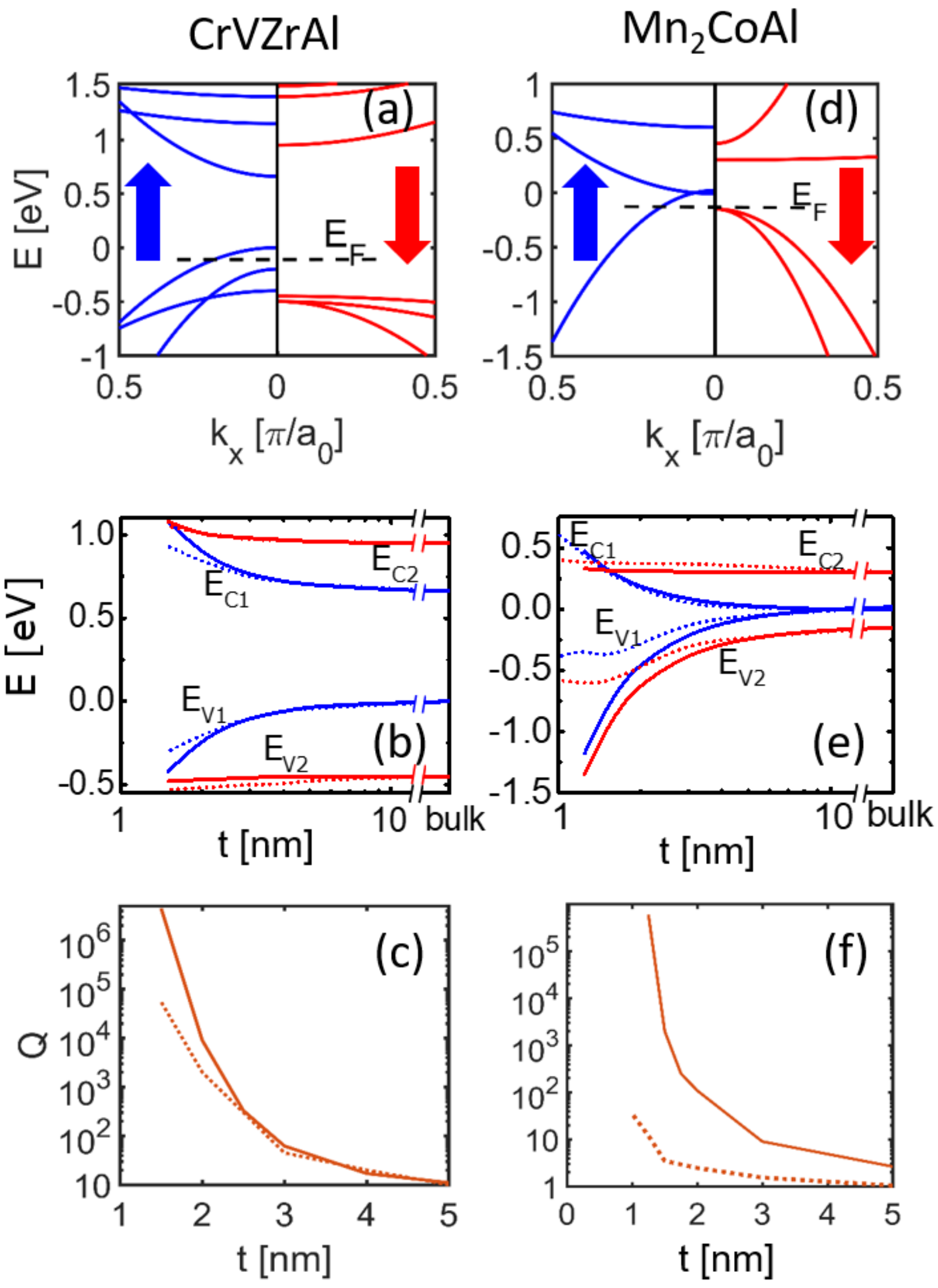

## Figure 2 caption:

The influence of quantum confinement on the bandstructure and *Q* factor in the Heuslers CrVZrAl (a,b,c) and $Mn_2CoAl$ (d,e,f) using parabolic band approximation and particle-in-a-box quantization. In (b), (c), (e) and (f), the dotted lines represent the data calculated considering numerical non-parabolic bands, as described in the main text. (a, d) Bulk bandstructures, and (b, e) shift of the band edges with confinement for CrVZrAl and $Mn_2CoAl$, respectively, versus film thickness. $E_{C1}$, $E_{C2}$, $E_{V1}$, $E_{V2}$ are the majority and minority conduction bands and valence bands edges, respectively. (c, f) The *Q* factor for the CrVZrAl and $Mn_2CoAl$ cases, respectively, for small layer thicknesses.

Figure 3:

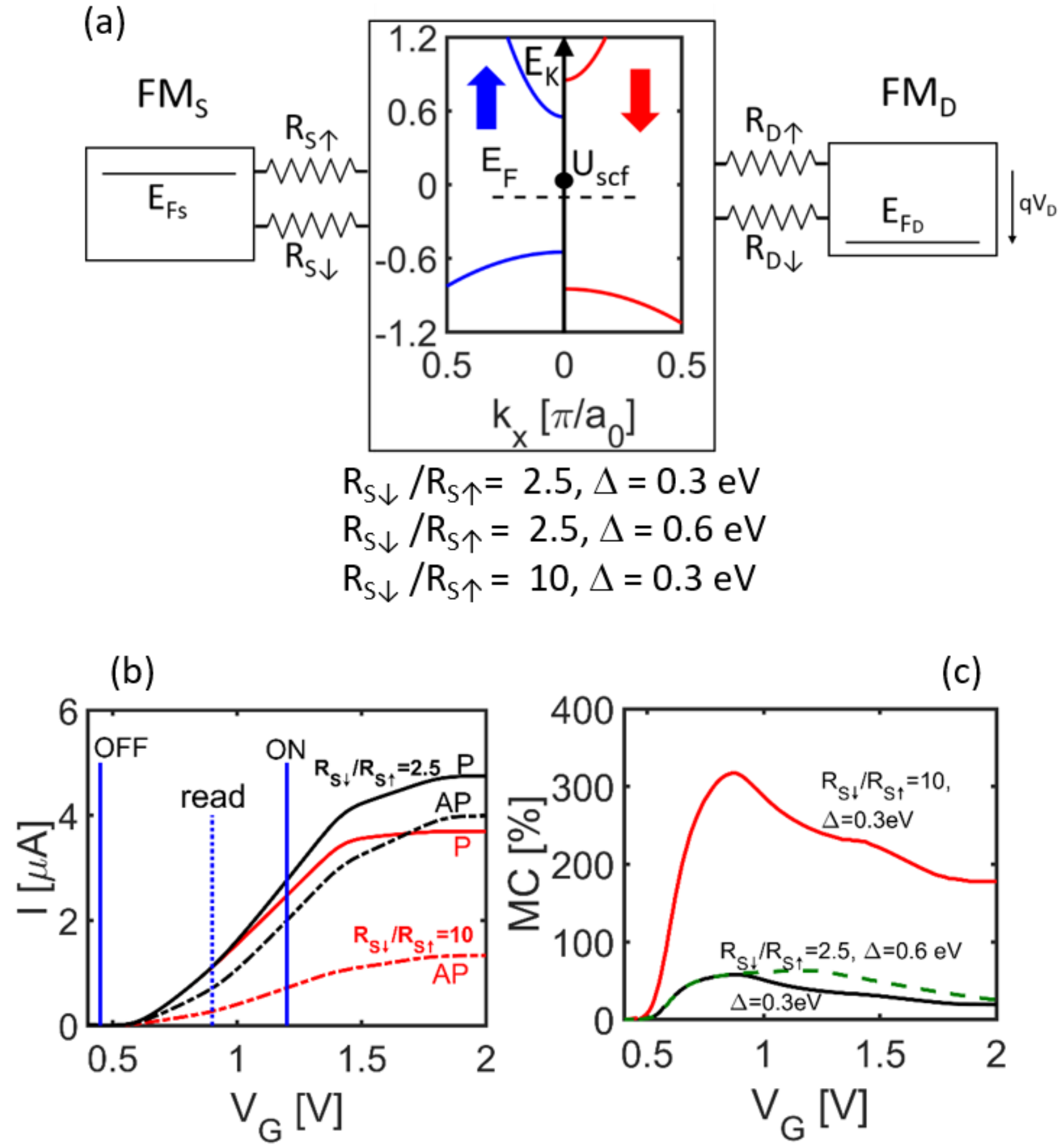

## Figure 3 caption:

The spin-FET device operation. (a) The spin-FET model with spin dependent series resistances introduced for the majority and minority carriers at the source/drain contacts. The values for the ratio of the parallel (P) and antiparallel (AP) cases are indicated with the $R_{\uparrow}$ value being $10^5$ Ω. The device is symmetric so that the resistances at source and drain have the same values. (b) Drain current versus gate voltage characteristics at $V_D$ = 0.75 V for Δ = 0.3 eV (a typical value for Heusler alloys) for two spin dependent resistance combinations. In **black lines** the case for $R_{\uparrow}=10^5$ Ω and $R_{\downarrow}=2.5\cdot10^5$ Ω is shown, which is as calculated in the text. In **red lines** the case for $R_{\uparrow}=10^5$ Ω and $R_{\downarrow}=10^6$ Ω is shown. Parallel (P) and antiparallel (AP) orientations are shown by the **solid** and **dash-dot lines**, respectively. The vertical solid blue lines show $V_G^{off}$ and $V_G^{on}$ for which high $I_{on}/I_{off}$ ratio is achived for a bias window $V_G = V_D$ = 0.75 V for both orientations ($I_{on}/I_{off}$ is ~ $10^4$ in all the cases, and is only slightly affected by the increased minority spin contact resistance). The vertical **dotted blue line** represents the ‘read’ gate bias $V_G^{read}$ for memory operation. (c) The magnetoconductance (MC) percentage as a function of $V_G$ for $V_D = 0.75$ V for three device parameter combinations as indicated, which shows separately the effect of Δ and $R_\uparrow / R_\downarrow$ on the MC.

Figure 4:

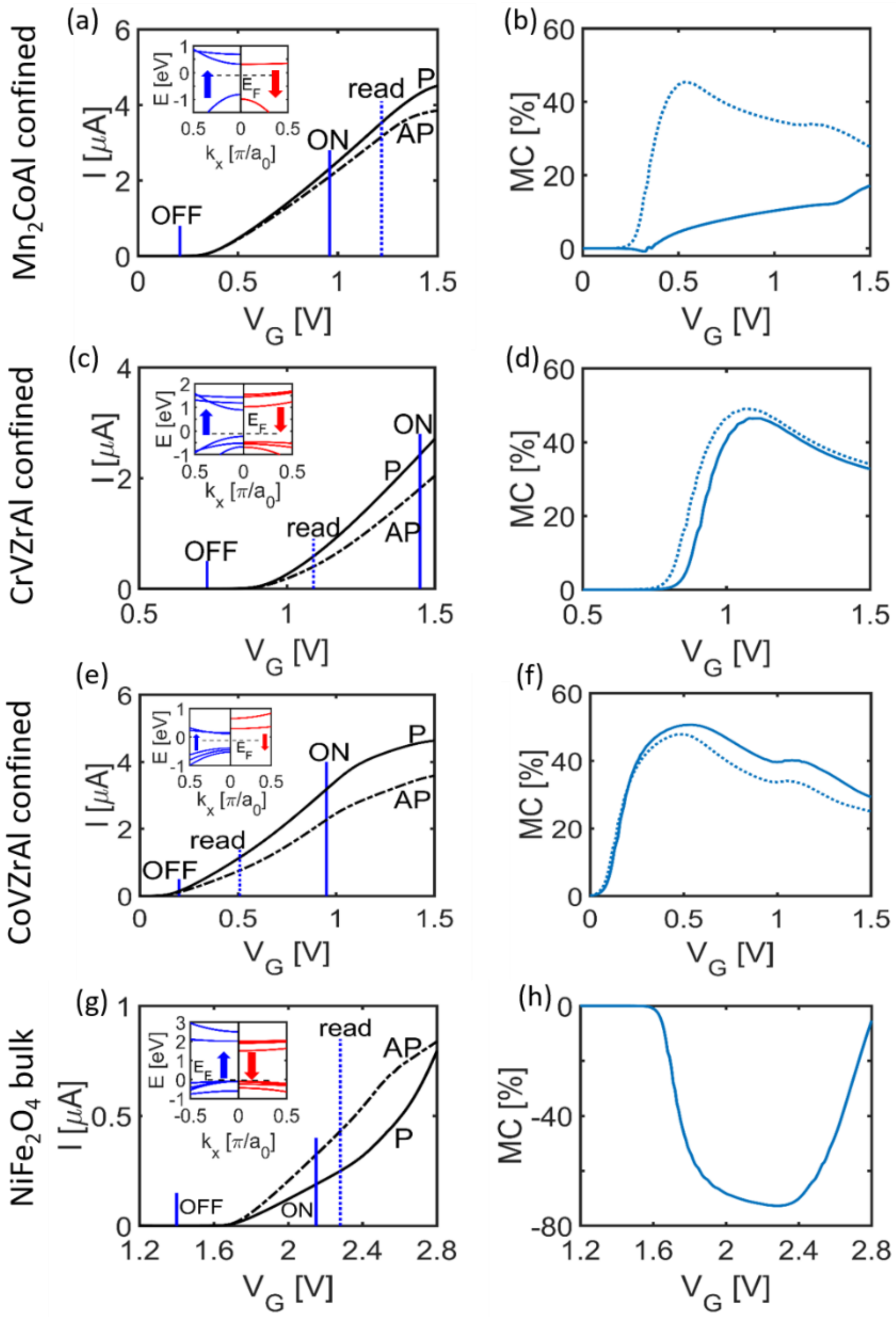

<u>Figure 4 caption:</u>

Drain current magnetoconductance (MC) percentage versus gate bias for real materials. (a) The $I_D$-$V_G$ for a $t$ = 1.5 nm narrow $Mn_2CoAl$ Heusler alloy channel with parallel (P) and anti-parallel (AP) current configurations shown by solid and dash-dot lines, respectively and noted. The $I_{on}/I_{off}$ ratio is ~ $10^3$ for both the P and AP cases. (b) MC versus gate bias for (a). The dotted line represents the MC when taking into account non-parabolicity effects, that increase the MC while the $I_D$-$V_G$ is only slightly affected (not shown). (c)-(d) Same as

(a) and (b) for a $t$ = 2 nm narrow CrVZrAl Heusler channel. The $I_{on}/I_{off}$ ratio is ~ $10^3$. (e)-(f) Same as (a)-(b) and (c)-(d) for a $t$ = 1.5 nm narrow CoVZrAl channel. The $I_{on}/I_{off}$ ratio is higher at ~ $10^4$. (g) The $I_D$-$V_G$ for the oxide $NiFe_2O_4$. (h) MC versus gate bias for (g). The $I_{on}/I_{off}$ ratio is ~ $10^6$ for both the P and AP cases, respectively. The corresponding bandstructures of the materials are shown in the insets, labelling the majority (blue) and minority (red) bands. In all cases $V_D$ = 0.75 V. In the $NiFe_2O_4$ case, the negative MC originates from the fact that the minority bands have a lower bandgap.

# Table 1

| **$Mn_2CoAl$** | $CB_1^{\uparrow}$ | $CB_2^{\uparrow}$ | $VB_1^{\uparrow}$ | $VB_2^{\uparrow}$ | $VB_3^{\uparrow}$ | $CB_1^{\downarrow}$ | $CB_2^{\downarrow}$ | $VB_1^{\downarrow}$ | $VB_2^{\downarrow}$ | $VB_3^{\downarrow}$ |
|---|---|---|---|---|---|---|---|---|---|---|
| $E_{offset}$(eV) | -0.01 | 0.6 | 0.02 | 0.02 | 0.02 | 0.3 | 0.45 | -0.14 | -0.14 | -0.14 |
| m* | 0.5 | 2.0 | 0.2 | 0.2 | 0.2 | 10.0 | 0.1 | 0.2 | 0.2 | 0.1 |

| **CrVZrAl** | $CB_1^{\uparrow}$ | $CB_2^{\uparrow}$ | $CB_3^{\uparrow}$ | $VB_1^{\uparrow}$ | $VB_2^{\uparrow}$ | $VB_3^{\uparrow}$ | $CB_1^{\downarrow}$ | $CB_2^{\downarrow}$ | $CB_3^{\downarrow}$ | $VB_1^{\downarrow}$ | $VB_2^{\downarrow}$ | $VB_3^{\downarrow}$ |
|---|---|---|---|---|---|---|---|---|---|---|---|---|
| $E_{offset}$(eV) | 0.66 | 1.15 | 1.4 | 0.0 | -0.2 | -0.4 | 0.95 | 1.4 | 1.5 | -0.45 | -0.5 | -0.5 |
| m* | 0.4 | 2.3 | 3.4 | 0.4 | 0.2 | 0.8 | 1.3 | 1.6 | 2.0 | 5.0 | 1.9 | 0.5 |

| **CoVZrAl** | $CB_1^{\uparrow}$ | $CB_2^{\uparrow}$ | $VB_1^{\uparrow}$ | $VB_2^{\uparrow}$ | $VB_3^{\uparrow}$ | $CB_1^{\downarrow}$ | $CB_2^{\downarrow}$ | $VB_1^{\downarrow}$ | $VB_2^{\downarrow}$ | $VB_3^{\downarrow}$ |
|---|---|---|---|---|---|---|---|---|---|---|
| $E_{offset}$(eV) | 0.0 | 0.1 | -0.25 | -0.25 | -0.25 | 0.25 | 0.55 | -1.0 | -1.0 | -1.0 |
| m* | 1.4 | 3.2 | 0.6 | 0.8 | 1.2 | 4.0 | 1.6 | 0.8 | 1.1 | 1.5 |

| **$NiFe_2O_4$** | $CB_1^{\uparrow}$ | $CB_2^{\uparrow}$ | $CB_3^{\uparrow}$ | $CB_4^{\uparrow}$ | $VB_1^{\uparrow}$ | $VB_2^{\uparrow}$ | $VB_3^{\uparrow}$ | $VB_4^{\uparrow}$ |
|---|---|---|---|---|---|---|---|---|
| $E_{offset}$(eV) | 2.0 | 2.0 | 2.0 | 2.5 | -0.1 | -0.1 | -0.1 | -0.6 |
| m* ($m_0$) | 2.0 | 1.6 | 1.1 | 0.3 | 0.3 | 0.4 | 1.6 | 0.8 |
| | $CB_1^{\downarrow}$ | $CB_2^{\downarrow}$ | $CB_3^{\downarrow}$ | $CB_4^{\downarrow}$ | $VB_1^{\downarrow}$ | $VB_2^{\downarrow}$ | $VB_3^{\downarrow}$ | $VB_4^{\downarrow}$ |
| $E_{offset}$(eV) | 1.5 | 1.9 | 1.9 | 2.0 | -0.1 | -0.14 | -0.25 | -0.45 |
| m* ($m_0$) | 1.1 | 1.9 | 2.8 | 3.3 | 1.1 | 0.8 | 1.9 | 0.7 |

## Table 1 caption:

Table 1 contains the bandstructure effective mass and band splitting data, extracted from DFT calculations in literature, used here for the real material simulations, $Mn_2CoAl$ (10 bands),[17] CrVZrAl (12 bands),[18] CoVZrAl (10 bands),[14] and $Ni_2FeO_4$ (16 bands).[19] The lattice parameters are 0.5798 nm, 0.641 nm, 0.626 nm and 0.833 nm for $Mn_2CoAl$, CrVZrAl, CoVZrAl and $NiFe_2O_4$, respectively. Energy gaps ($E_G$) are -0.03 eV, 0.66 eV, 0.25 eV and 1.6 eV for $Mn_2CoAl$, CrVZrAl, CoVZrAl and $NiFe_2O_4$, respectively. The effective masses $m^*$ are extracted using a parabolic band approximation in units of electron rest mass $m_0$. $E_{offset}$ is the energy position of each band's minimum (in the case of CBs) and of each band's maximum (in the case of VBs) with respect to the zero as denoted in the DFT data. Each band is numbered in a progressive order starting from the one closest to the $E_F$ = -0.1 eV that we set. These are the bands closer to $E_F$, which are the ones involved in transport at the considered biases. Upward (downward) arrows indicate the majority (minority) spin direction.

# Appendix

We show here the calculations of the interfacial voltage drop due to spin flip at the interface between two ferromagnets A and B for the parallel (P) and antiparallel (AP) situations. This voltage drop is responsible for the spin dependent contact resistance. These calculations consider two ferromagnets in contact, but can be extended to describe the interface between a ferromagnet and a non-ferromagnetic material which, however, carries a spin polarized current. This can be the situation where a spin polarized current is injected in a semiconductor in steady-state conditions.

We follow the approach proposed in Ref. [36] and in the Appendix C therein. We place the interface at $x = 0$ so $x < 0$ for A and $x > 0$ for B. The 'up' arrows denote majority spins and the 'down' arrows the minority ones. Note that we use the letter $\beta$ to indicate the spin polarization SP for a more concise notation: SP $\equiv \beta$.

In general we have:

$$J_{\uparrow}^{\mathrm{A}} = (1+\beta_{\mathrm{A}})\frac{J}{2} - \frac{1}{2e\rho_{\mathrm{A}} l_s^{\mathrm{A}}} K_2^{\mathrm{A}} e^{x/l_s^{\mathrm{A}}} \tag{A.1a}$$

$$J_{\downarrow}^{\mathrm{A}} = (1-\beta_{\mathrm{A}})\frac{J}{2} + \frac{1}{2e\rho_{\mathrm{A}} l_s^{\mathrm{A}}} K_2^{\mathrm{A}} e^{x/l_s^{\mathrm{A}}} \tag{A.1b}$$

$$J_{\uparrow}^{\mathrm{B}} = (1+\beta_{\mathrm{B}})\frac{J}{2} - \frac{1}{2e\rho_{\mathrm{B}} l_s^{\mathrm{B}}} K_3^{\mathrm{B}} e^{-x/l_s^{\mathrm{B}}} \tag{A.1c}$$

$$J_{\downarrow}^{\mathrm{B}} = (1-\beta_{\mathrm{B}})\frac{J}{2} + \frac{1}{2e\rho_{\mathrm{B}} l_s^{\mathrm{B}}} K_3^{\mathrm{B}} e^{-x/l_s^{\mathrm{B}}} \tag{A.1d}$$

In the above equations the signs invert when magnetization switches.[36] $K_2$ and $K_3$ are constants to be determined and the significance of the other symbols is reported in the main text. Note that $x < 0$ for A and $x > 0$ for B.

**Figure A1** sketches the current scheme at the interface for the case when the A and B ferromagnets are oriented in parallel (top) and antiparallel (bottom). We always consider $\beta_{\mathrm{A}} > \beta_{\mathrm{B}}$. $x$ is the spatial coordinate of the unidirectional current flow, and the junction interface is located at $x = 0$. $J_{\uparrow(\downarrow)}^{\mathrm{A(B)}}$ indicates the current for the majority (↑) or minority (↓) spin direction in materials A and B. The solid lines represent the current values so that the overall flowing current *J*, given by the sum of the two components, is preserved in space. The dotted line represents the value of half of the total current. Far from the interface, the deviation from such a value of the majority/minority spin currents depends on the value of $\beta$. $l_S^{A(B)}$ is the spin diffusion length in A or B as noted and represents the characteristic length of the exponential decay of the current.

In the parallel alignment case, because $\beta_A > \beta_B$, in A the two spin currents are separated more than in B. Hence, at the interface the majority spin current has to decrease and the minority spin current has to increase, thus, spin flip events are necessary. The higher the difference in $\beta$, the stronger the spin flip and the higher the voltage drop. If $\beta_A = \beta_B$, the voltage drop would be zero and the current lines would be straight across the junction.

In the antiparallel case, as B is oriented in the antiparallel configuration with respect to A, the higher current component is for the minority spins (according to the notation in A). Thus, the highest current component in A turns into the lowest current component in B, the two spin components cross each other, and the current spin polarization inverts its sign. This happens at the interface where the net spin polarization is zero. Because the current value is always *J*, at the interface each component has *J/2* value. In this case, the spin flip events are stronger as they are responsible for the inversion of the spin polarization of the current, and the related voltage drop is higher leading to a higher junction resistance.

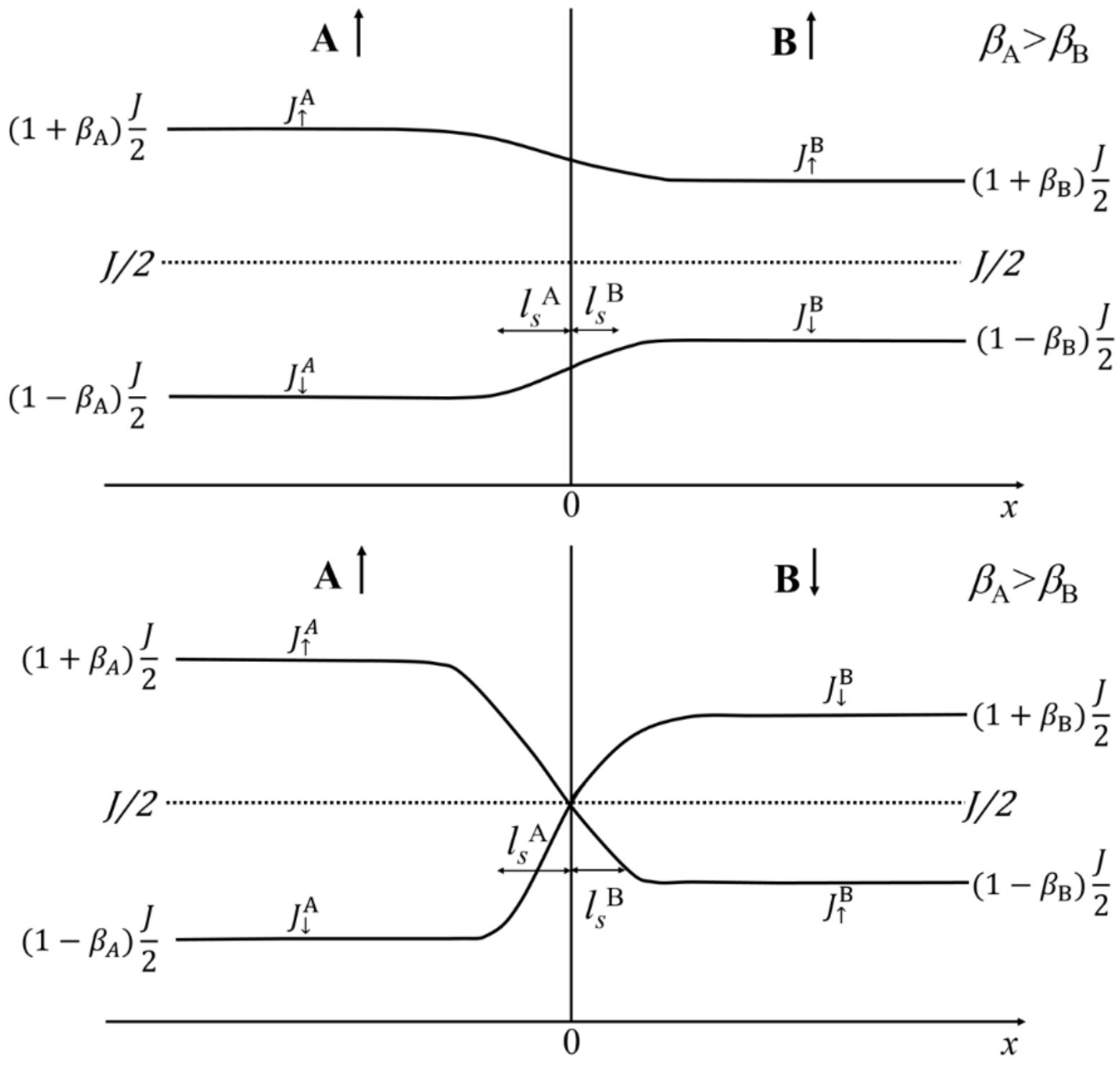

Figure A1: The current scheme at the interface for the case when the A and B ferromagnets are oriented in parallel (top) and antiparallel (bottom). *x* is the spatial coordinate of the unidirectional current flow, and the junction interface is located at $x = 0$. $l_S^{A(B)}$ is the spin diffusion length in A or B. $J_{\uparrow(\downarrow)}^{A(B)}$ indicates the current for the majority (↑) or minority (↓) spin direction in A and B. The solid lines represent the current values, the dotted line represents the value of half of the total current. Far from the interface, the deviation from such a value of the majority/minority spin currents depends on the value of $\beta$.

For the **parallel** case, we assume that each current component at the interface takes a value that is the average of the value taken far from the interface:

$$J^{A}_{\uparrow(x=0)} = J^{B}_{\downarrow(x=0)} = \frac{\lim_{x\to-\infty} J^{A}_{\uparrow} + \lim_{x\to\infty} J^{B}_{\uparrow}}{2} = \frac{(1+\beta_A)\frac{J}{2}+(1+\beta_B)\frac{J}{2}}{2} = \frac{J}{2}\frac{2+\beta_A+\beta_B}{2} \equiv \frac{J}{2}\beta_{+}^{eff} \qquad \text{(A.2a)}$$

$$J^{A}_{\downarrow(x=0)} = J^{B}_{\downarrow(x=0)} = \frac{\lim_{x\to-\infty} J^{A}_{\downarrow} + \lim_{x\to\infty} J^{B}_{\downarrow}}{2} = \frac{(1-\beta_A)\frac{J}{2}+(1-\beta_B)\frac{J}{2}}{2} = \frac{J}{2}\frac{2-(\beta_A+\beta_B)}{2} \equiv \frac{J}{2}\beta_{-}^{eff} \qquad \text{(A.2b)}$$

Then, at the interface ($x$=0), for the A side:

$$(1-\beta_A)\frac{J}{2} + \frac{1}{2e\rho_A l_s^A} K_2^A = \frac{J}{2}\beta_{-}^{eff} \rightarrow K_2^A = \frac{J}{2} e\rho_A l_s^A (\beta_{-}^{eff} - 1 + \beta_A) = \frac{J}{2} e\rho_A l_s^A \Delta\beta \qquad \text{(A.3a)}$$

and, for the B side:

$$(1-\beta_{\mathrm{B}})\frac{J}{2}+\frac{1}{2e\rho_{\mathrm{A}}l_s^{\mathrm{A}}}K_3^{\mathrm{B}}=\frac{J}{2}\beta_-^{eff}\rightarrow K_3^{\mathrm{B}}=\frac{J}{2}e\rho_{\mathrm{B}}l_s^{\mathrm{B}}(\beta_-^{eff}-1+\beta_{\mathrm{B}})=-\frac{J}{2}e\rho_{\mathrm{A}}l_s^{\mathrm{A}}\Delta\beta \quad \text{(A.3b)}$$

where $\Delta\beta=\beta_{\mathrm{A}}-\beta_{\mathrm{B}}$ with $\Delta\beta>0$.

For the **antiparallel** case, at the interface, where the current components invert, each of them takes the same value of *J/2*, in analogy with Ref. [36]:

$$J_{\uparrow(x=0)}=J_{\downarrow(x=0)}=\frac{J}{2} \quad \text{(A.4)}$$

Then, at the interface (*x*=0), for the A side:

$$(1-\beta_{\mathrm{A}})\frac{J}{2}+\frac{1}{2e\rho_{\mathrm{A}}l_s^{\mathrm{A}}}K_2^{\mathrm{A}}=\frac{J}{2}\rightarrow K_2^{\mathrm{A}}=Je\rho_{\mathrm{A}}l_s^{\mathrm{A}}\beta_{\mathrm{A}} \quad \text{(A.5a)}$$

and for the B side:

$$(1+\beta_{\mathrm{B}})\frac{J}{2}+\frac{1}{2e\rho_{\mathrm{A}}l_s^{\mathrm{A}}}K_3^{\mathrm{B}}=\frac{J}{2}\rightarrow K_3^{\mathrm{B}}=-Je\rho_{\mathrm{B}}l_s^{\mathrm{B}}\beta_{\mathrm{B}} \quad \text{(A.5b)}$$

In order to calculate the voltage drops at the interface we need to obtain the *F(x)* functions, i.e. the spatial gradient of the spin dependent electrochemical potential divided by the electron charge, which have the dimension of an electric field, expressed as in the appendix C of Ref. [36]:

$$F^{\mathrm{A}}(x)=(1-\beta_{\mathrm{A}}^2)\rho_{\mathrm{A}}J+\frac{\beta_{\mathrm{A}}}{el_s^{\mathrm{A}}}\left[K_2^{\mathrm{A}}exp\left(\frac{x}{l_s^{\mathrm{A}}}\right)\right] \quad \text{(A.6a)}$$

$$F^{\mathrm{B}}(x)=(1-\beta_{\mathrm{B}}^2)\rho_{\mathrm{B}}J-\frac{\beta_{\mathrm{B}}}{el_s^{\mathrm{B}}}\left[K_3^{\mathrm{B}}exp\left(-\frac{x}{l_s^{\mathrm{B}}}\right)\right] \quad \text{(A.6b)}$$

We can now calculate $F(x) - E_0$ that is the argument of the integrals that give the interfacial voltage drop $\Delta V_I^{P/AP}$,[36] where ${E_0}^{\mathrm{A(B)}} = (1-\beta_{\mathrm{A(B)}}^2)\rho_{\mathrm{A(B)}}J$ is the unperturbed electric field (i.e., far from the interface).

$$\Delta V_I = \int_{-\infty}^{+\infty}[F(x) - E_0]d\,x \qquad \text{(A.7)}$$

For the **parallel** case:

$$F^{\mathrm{A}}(x) - {E_0}^{\mathrm{A}} = \frac{J}{2}\rho_{\mathrm{A}}\beta_{\mathrm{A}}\Delta\beta exp\left(\frac{x}{l_s^{\mathrm{A}}}\right) \qquad \text{(A.8a)}$$

$$F^{\mathrm{B}}(x) - {E_0}^{\mathrm{B}} = \frac{J}{2}\rho_{\mathrm{B}}\beta_{\mathrm{B}}\Delta\beta exp\left(-\frac{x}{l_s^{\mathrm{B}}}\right) \qquad \text{(A8.b)}$$

For the **antiparallel** case:

$$F^{\mathrm{A}}(x) - {E_0}^{\mathrm{A}} = J\rho_{\mathrm{A}}\beta_{\mathrm{A}}^2\, exp\left(\frac{x}{l_s^{\mathrm{A}}}\right) \qquad \text{(A.9a)}$$

$$\mathrm{F}^{\mathrm{B}}(x) - {\mathrm{E}_0}^{\mathrm{B}} = J\rho_{\mathrm{B}}\beta_{\mathrm{B}}^2\, exp\left(-\frac{x}{l_s^{\mathrm{B}}}\right) \qquad \text{(A.9b)}$$

Hence:

$$\Delta V_I^{\mathrm{P}} = \int_{-\infty}^{0}\frac{J}{2}\rho_{\mathrm{A}}\beta_{\mathrm{A}}\Delta\beta exp\left(\frac{x}{l_s^{\mathrm{A}}}\right)d\,x + \int_{0}^{+\infty}\frac{J}{2}\rho_{\mathrm{B}}\beta_{\mathrm{B}}\Delta\beta exp\left(-\frac{x}{l_s^{\mathrm{B}}}\right)dx =$$

$$= \frac{J}{2}(\beta_{\mathrm{A}}\Delta\beta\rho^{\mathrm{A}}l_s^{\mathrm{A}} + \beta_{\mathrm{B}}\Delta\beta\rho^{\mathrm{B}}l_s^{\mathrm{B}}) \qquad \text{(A.10)}$$

and

$$\Delta V_I^{\mathrm{AP}} = \int_{-\infty}^{0}\rho^{\mathrm{A}}J\,\beta_{\mathrm{A}}^2\exp\left(\frac{x}{l_s^{\mathrm{A}}}\right)dx + \int_{0}^{+\infty}\rho^{\mathrm{B}}J\beta_{\mathrm{B}}^2\exp\left(-\frac{x}{l_s^{\mathrm{B}}}\right)dx = J(\beta_{\mathrm{A}}^2\rho^{\mathrm{A}}l_s^{\mathrm{A}} + \beta_{\mathrm{B}}^2\rho^{\mathrm{B}}l_s^{\mathrm{B}}). \qquad \text{(A.11)}$$

Finally, the interface resistance per unit area can be estimated from $\Delta V_I/J$.